%% file: part1_v3_noblue.tex
\documentclass[journal,subeqn,caption]{IEEEtran}
\usepackage{mathtools}
\usepackage{algorithm}
\usepackage{algpseudocode}
\usepackage{hhline}
\usepackage{pdflscape}
\usepackage{balance}

\usepackage{cite}
\usepackage{float}
\usepackage{multicol}
\usepackage{multirow}
\usepackage{balance}
\usepackage{graphicx}
\usepackage{amssymb}
\usepackage{cuted}

\usepackage{rotating,booktabs,dcolumn}
\usepackage[flushleft]{threeparttable}

\DeclarePairedDelimiter\abs{\lvert}{\rvert}%
\DeclarePairedDelimiter\norm{\lVert}{\rVert}%

\makeatletter
\let\oldabs\abs
\def\abs{\@ifstar{\oldabs}{\oldabs*}}
\let\oldnorm\norm
\def\norm{\@ifstar{\oldnorm}{\oldnorm*}}
\makeatother

\usepackage{array}
\newcolumntype{P}[1]{>{\centering\arraybackslash}p{#1}}
\newcolumntype{M}[1]{>{\centering\arraybackslash}m{#1}}

\usepackage[hidelinks]{hyperref}
\usepackage{graphicx}
\usepackage{amsfonts}
\usepackage{amsmath}
\usepackage{mathabx}
\usepackage{bm}
\usepackage{color}
\usepackage{epstopdf}
\usepackage{mathtools, nccmath}
\usepackage{multirow}
\usepackage{bbm}
\usepackage{marginnote}
\usepackage{tikz}
\usetikzlibrary{shapes,arrows,chains}
\usepackage{enumitem}
\usepackage{comment}
\usepackage{eurosym}
 \usepackage{epstopdf}
\usetikzlibrary{arrows,decorations.pathmorphing,backgrounds,fit,positioning,shapes.symbols,chains}
\usepackage{caption}
\usepackage{colortbl}
\usepackage{float}
\usepackage{xcolor}
\usepackage[caption=false]{subfig}
\usepackage{tabularx}

\definecolor{Gray}{gray}{0.9}
\definecolor{LightCyan}{rgb}{0.88,1,1}

\newcolumntype{a}{>{\columncolor{Gray}}c}
\newcolumntype{b}{>{\columncolor{white}}c}

\makeatletter
\newcases{Rcases}{$\mskip\thickmuskip$}{%
  \hfil$\m@th{##}$}{$\m@th{##}$\hfil}{.}{\rbrace}
\makeatother
\allowdisplaybreaks

 \ifCLASSINFOpdf
\else
\fi
\newenvironment{ldescription}[1]
  {\begin{list}{}%
   {\renewcommand\makelabel[1]{##1\hfill}%
   \settowidth\labelwidth{\makelabel{#1}}%
   \setlength\leftmargin{\labelwidth}
   \addtolength\leftmargin{\labelsep}}}
  {\end{list}}

\begin{document}
\newcommand{\AF}{E}
\newcommand{\AT}{E^\mathrm{R}}
\newcommand{\A}{\AF \!\cup \!\AT}
\newcommand{\BP}{N^{\mathrm{P}}}
\newcommand{\BL}{L_i}
\newcommand{\BS}{S_i}
\newcommand{\G}{G}
\newcommand{\BG}{G_i}
\newcommand{\RB}{R}

\newcommand{\BPR}{N^{\mathrm{PR}}}

\newcommand{\BB}{\beta}

\newcommand{\Xiul}{\Xi^{\mathrm{ul}}}
\newcommand{\Xill}{\Xi^{\mathrm{ll}}}
\newcommand{\Xir}{\Xi^{\mathrm{r}}}
\newcommand{\Xidu}{\Xi^{\mathrm{du}}}

\renewcommand{\wp}{op}
\newcommand{\es}{es}

\newcommand{\CC}{\bm{\ddot{c}}_{k}}
\newcommand{\C}{\bm{\dot{c}}_{k}}
\newcommand{\cc}{\bm{c}_{k}}
\newcommand{\Pgen}{P^{\mathrm{g}}_{t,k}}
\newcommand{\Qgen}{Q^{\mathrm{g}}_{t,k}}
\newcommand{\Pgenm}{P^{\mathrm{g}}_{t-1,k}}
\newcommand{\Qgenm}{Q^{\mathrm{g}}_{t-1,k}}
\renewcommand{\P}{P_{t,e,i,j}}
\newcommand{\Q}{Q_{t,e,i,j}}
\newcommand{\Sline}{\bm{\overline{S}}_{e}}
\newcommand{\dP}{\bm{P}^{\mathbf{d}}_{t,l}}
\newcommand{\dQ}{\bm{Q}^{\mathbf{d}}_{t,l}}
\newcommand{\Vtn}{\bm{V}^{\mathbf{\wp}}_{t,i}}
\newcommand{\Vtm}{\bm{V}^{\mathbf{\wp}}_{t,j}}
\newcommand{\fitn}{\bm{\theta}^{\mathbf{\wp}}_{t,i}}
\newcommand{\fitm}{\bm{\theta}^{\mathbf{\wp}}_{t,j}}
\newcommand{\dVn}{V^{\Delta}_{t,i}}
\newcommand{\dVm}{V^{\Delta}_{t,j}}
\newcommand{\dfin}{\theta^{\Delta}_{t,i}}
\newcommand{\dfim}{\theta^{\Delta}_{t,j}}

\newcommand{\Va}{\widecheck{V}_{t,e}}

\newcommand{\cosf}{\widehat{cos}_{t,i,j}}
\newcommand{\cost}{\widehat{cos}_{t,j,i}}

\newcommand{\gl}{\bm{g}_{e}}
\newcommand{\gf}{\bm{g}^{\mathbf{fr}}_{e}}
\newcommand{\gt}{\bm{g}^{\mathbf{to}}_{e}}

\newcommand{\bl}{\bm{b}_{e}}
\renewcommand{\bf}{\bm{b}^{\mathbf{fr}}_{e}}
\newcommand{\bt}{\bm{b}^{\mathbf{to}}_{e}}

\newcommand{\gs}{\bm{g}^{\mathbf{sh}}_{s}}
\newcommand{\bs}{\bm{b}^{\mathbf{sh}}_{s}}

\newcommand{\tap}{\bm{\tau}_{e}}
\newcommand{\shift}{\bm{\sigma}_{e}}

\newcommand{\Pmax}{\overline{\bm{P}}^{\mathbf{g}}_k}
\newcommand{\Pmin}{\underline{\bm{P}}^{\mathbf{g}}_k}
\newcommand{\Qmax}{\overline{\bm{Q}}^{\mathbf{g}}_k}
\newcommand{\Qmin}{\underline{\bm{Q}}^{\mathbf{g}}_k}
\newcommand{\Vmax}{\overline{\bm{V}}_{i}}
\newcommand{\Vmin}{\underline{\bm{V}}_{i}}

\newcommand{\cospsinij}{\bm{cps}_{t,e,i,j}}
\newcommand{\cospsinji}{\bm{cps}_{t,e,j,i}}
\newcommand{\cosmsinij}{\bm{cms}_{t,e,i,j}}
\newcommand{\cosmsinji}{\bm{cms}_{t,e,j,i}}

\newcommand{\vpJedan}{\bm{p}_{1,t,e,i,j}}
\newcommand{\vpDva}{\bm{p}_{2,t,e}}
\newcommand{\vpTri}{\bm{p}_{3,t,e,i,j}}

\newcommand{\condv}{\bm{\Lambda}_{t,e}}
\newcommand{\condc}{\bm{\Gamma}_{t,i,j}}
\newcommand{\condcji}{\bm{\Gamma}_{t,j,i}}
\newcommand{\conds}{\bm{\Phi}_{t,e,i,j}}

\newcommand{\SoEmax}{\overline{\bm{SoE}}}
\newcommand{\etach}{\bm{\eta}^{\mathbf{ch}}}
\newcommand{\etadis}{\bm{\eta}^{\mathbf{dis}}}
\newcommand{\qchmax}{\bm{\overline{s}}^{\mathbf{\es}}}
\newcommand{\qdismax}{\bm{\overline{s}}^{\mathbf{\es}}}

\newcommand{\SoE}{SoE_{t}}
\newcommand{\SoEm}{SoE_{t-1}}
\newcommand{\qbat}{p^{\mathrm{\es}}_{t}}
\newcommand{\qch}{p^{\mathrm{ch}}_{t}}
\newcommand{\qdis}{p^{\mathrm{dis}}_{t}}
\newcommand{\qqbat}{q^{\mathrm{\es}}_{t}}
\newcommand{\qqch}{q^{\mathrm{ch}}_{t}}
\newcommand{\qqdis}{q^{\mathrm{dis}}_{t}}

\newcommand{\xch}{x^{\mathrm{p}}_{t}}
\newcommand{\xxch}{x^{\mathrm{q}}_{t}}

\newcommand{\lambdaJedan}{\lambda_{1,t,i}}
\newcommand{\lambdaJedanDn}{\underline{\bm{\lambda}}_{1,t,i}}
\newcommand{\lambdaJedanUp}{\overline{\bm{\lambda}}_{1,t,i}}

\newcommand{\lambdaDva}{\lambda_{2,t,i}}
\newcommand{\lambdaDvaDn}{\underline{\bm{\lambda}}_{2,t,i}}
\newcommand{\lambdaDvaUp}{\overline{\bm{\lambda}}_{2,t,i}}

\newcommand{\lambdaTri}{\lambda_{3,t,e,i,j}}
\newcommand{\lambdaTriji}{\lambda_{3,t,e,j,i}}
\newcommand{\lambdaCetiri}{\lambda_{4,t,e,i,j}}
\newcommand{\lambdaCetiriji}{\lambda_{4,t,e,j,i}}
\newcommand{\lambdaPet}{\lambda_{5,t,e,i,j}}
\newcommand{\lambdaPetji}{\lambda_{5,t,e,j,i}}
\newcommand{\lambdaSest}{\lambda_{6,t,e,i,j}}
\newcommand{\lambdaSestji}{\lambda_{6,t,e,j,i}}
\newcommand{\lambdaSedam}{\lambda_{16,t,i}}

\newcommand{\lambdaJedanaest}{\lambda_{14,t,e,i,j}}
\newcommand{\lambdaDvanaest}{\lambda_{15,t,e,i,j}}
\newcommand{\lambdaTrinaest}{\mu_{5,t,e,i,j}}

\newcommand{\lambdaCetrnaest}{\lambda_{11,t,i,j}}
\newcommand{\lambdaPetnaest}{\lambda_{12,t,i,j}}
\newcommand{\lambdaSesnaest}{\mu_{2,t,i,j}}
\newcommand{\lambdaCetrnaestji}{\lambda_{11,t,j,i}}

\newcommand{\lambdaSedamnaest}{\lambda_{7,t,e,i,j}}
\newcommand{\lambdaOsamnaest}{\lambda_{8,t,e,i,j}}
\newcommand{\lambdaOsamnaestji}{\lambda_{8,t,e,j,i}}
\newcommand{\lambdaDevetnaest}{\lambda_{9,t,e,i,j}}
\newcommand{\lambdaDvadeset}{\mu_{1,t,e,i,j}}

\newcommand{\lambdaDvadesetjedan}{\lambda_{10,t,e,i,j}}
\newcommand{\lambdaDvadesetdva}{\lambda_{13,t,i,j}}

\newcommand{\muJedanDn}{\underline{\mu}_{3,t,k}}
\newcommand{\muJedanUp}{\overline{\mu}_{3,t,k}}
\newcommand{\muDvaDn}{\underline{\mu}_{4,t,k}}
\newcommand{\muDvaUp}{\overline{\mu}_{4,t,k}}
\newcommand{\muTriDn}{\underline{\mu}_{6,t,i}}
\newcommand{\muTriUp}{\overline{\mu}_{6,t,i}}

\newcommand{\cvJedan}{f_{1,t,i,j}}
\newcommand{\cvDva}{f_{2,t,i,j}}
\newcommand{\cvTri}{f_{0,t,i,j}}

\newcommand{\vvJedan}{w_{1,t,e,i,j}}
\newcommand{\vvDva}{w_{2,t,e,i,j}}
\newcommand{\vvTri}{w_{3,t,e,i,j}}
\newcommand{\vvCetiri}{w_{0,t,e,i,j}}

\newcommand{\Omegad}{\Omega^{\mathrm{d}}}
\newcommand{\Omegap}{\Omega^{\mathrm{p}}}

\newcommand{\xnula}{x_{0}}
\newcommand{\xjedan}{x_{1}}
\newcommand{\xdva}{x_{2}}
\newcommand{\xtri}{x_{3}}
\newcommand{\xcrta}{\overline{x}}
\newcommand{\x}{x}

\newcommand{\ynula}{y_{0}}
\newcommand{\yjedan}{y_{1}}
\newcommand{\ydva}{y_{2}}
\newcommand{\ytri}{y_{3}}
\newcommand{\ycrta}{\overline{y}}
\newcommand{\y}{y}

\newcommand{\eps}{\bm{\epsilon}}
\newcommand{\smf}{F}
\newcommand{\psin}{\psi_{n}}
\newcommand{\psijedan}{\psi_{1}}
\newcommand{\psidva}{\psi_{2}}
\newcommand{\un}{u_{n}}
\newcommand{\ujedan}{u_{1}}
\newcommand{\udva}{u_{2}}

\newcommand{\wbat}{w^{\mathrm{\es}}_{t}}
\newcommand{\wch}{w^{\mathrm{ch}}_{t}}
\newcommand{\wdis}{w^{\mathrm{dis}}_{t}}

\newcommand{\ybat}{y^{\mathrm{\es}}_{t}}

\newcommand{\ib}{b}
\newcommand{\iu}{u}
\newcommand{\xbin}{x^{\mathrm{bin}}_{t,\ib}}

\newcommand{\US}{U}
\newcommand{\BSS}{B}
\newcommand{\wbe}{w^{\mathrm{be}}_{t,\ib}}
\newcommand{\wue}{w^{\mathrm{ue}}_{t,\iu}}
\newcommand{\xun}{x^{\mathrm{ue}}_{t,\iu}}

\newcommand{\pen}{\bm{\pi}}

%\bm{sc}_{k}

%Bi-level cluster model for optimal Battery Swapping Station operation

\title{Solving Bilevel AC OPF Problems by Smoothing the Complementary Conditions -- Part I: Model Description and the Algorithm}
\author{K. \v{S}epetanc, \textit{Student Member}, \textit{IEEE}, H. Pand\v{z}i\'c, \textit{Senior Member}, \textit{IEEE} and T. Capuder, \textit{Member}, \textit{IEEE}
%\thanks{The authors are with the University of Zagreb Faculty of Electrical Engineering and Computing (emails: karlo.sepetanc@fer.hr; hrvoje.pandzic@fer.hr). This work is a result of project Electric Vehicle BAttery Swapping Station - EVBASS, funded by the Croatian Science Foundation under grant IP-2014-09-3517.}
%
%
%\thanks{Manuscript received October 19, 2015; revised March 1, 2016.
%}
\thanks{The authors are with the Innovation Centre Nikola Tesla (ICENT) and the University of Zagreb Faculty of Electrical Engineering and Computing (e-mails: karlo.sepetanc@fer.hr; hrvoje.pandzic@fer.hr; tomislav.capuder@fer.hr). Employment of Karlo \v{S}epetanc is funded by the Croatian Science Foundation under programme DOK-2018-09. The research leading to these results has received funding from the European Union’s Horizon 2020 research and innovation programme under grant agreement No 864298 (project ATTEST). The sole responsibility for the content of this document lies with the authors. It does not necessarily reflect the opinion of the Innovation and Networks Executive Agency (INEA) or the European Commission (EC). INEA or the EC are not responsible for any use that may be made of the information contained therein.}
}
\maketitle

\begin{abstract}
%The low carbon transformation of the power systems is bringing changes to the known operational practices. The importance of precise modelling has never been more emphasized as the variabilities and uncertainties in the systems are increasing and are more often pushing technical limits to their boundaries. 
The existing research on market price-affecting agents, i.e. price makers, neglects or simplifies the nature of AC power flows in the power system as it predominantly relies on DC power flows. This paper proposes a novel bilevel formulation based on the smoothing technique, where any price-affecting strategic player can be modelled in the upper level, while the market clearing problem in the lower level uses convex quadratic transmission AC optimal power flow (AC OPF), with the goal of achieving accuracy close to the one of the exact nonlinear formulations. Achieving convexity in the lower level is the foundation for bilevel modeling since traditional single-level reduction techniques do not hold for nonconvex models. The bilevel market participation problem with the AC OPF formulation in the lower level is transformed into a single-level problem and solved using multiple techniques such as the primal-dual counterpart, the strong duality theorem, the McCormick envelopes, the complementary slackness, the penalty factor, the interaction discretization as well as the proposed smoothing techniques.  

Due to an extensive amount of information and descriptions, the overall work is presented as a two-part paper. This first part provides a literature overview, positions the work and presents the model and the solution algorithm, while the solution techniques and case studies are provided in the accompanying paper. %The superiority of the proposed model and smoothing techniques is demonstrated in terms of accuracy and computational tractability over several scenarios, multiple networks of different sizes and different OPF models. The results show that the proposed approach outperforms all other options in both metrics by a significant margin. This is especially noticeable in the metric of accuracy where the other methods can over/underestimate the bidding entity's profit by up to 13\% as opposed to 0.16\% with the proposed model.

\end{abstract}
%and iterative fast convergence
%The first presented case study is an optimal power flow that demonstrates accuracy of the proposed model and its iterative convergence

\section*{Nomenclature}
\subsection{Abbreviations}
\begin{ldescription}{$xxxxx$}
\item [OPF] Optimal power flow.
\item [SOC] Second-order cone.
\item [SOCP] Second-order cone programming.
\item [QC] Quadraticaly constrained.
\item [SDP] Semidefinite programming.
\item [IV] Current-voltage.
\item [LPAC] Linear programming AC.
\item [QPAC] Quadratic programming AC.
\item [ES] Energy storage.
\item [CPSOTA] Convex polar second-order Taylor approximation.
\item [KKT] Karush–Kuhn–Tucker.
\item [NLP] Nonlinear programming.
\item [OP] Operating point.
\item [NC] Nonconvex.
\item [LL] Lower level.
%\item [SoE] State-of-energy.
\end{ldescription}

\subsection{Sets and Indices}
\begin{ldescription}{$xxxxx$}
\item [$N$] Set of buses, indexed by $i$ and $j$.
\item [$\BB$] ES's bus location singleton, indexed by $i$.
\item [$\RB$] Singleton set containing reference bus, indexed by $i$.
\item [$\AF, \AT$] Tuple set of branches, forward and reverse orientation, indexed by $(e,i,j)$.
\item [$\BP,\BPR$] Tuple set of paired buses aligned with branch $\AF$ and $\AT$ orientations, indexed by $(i,j)$.
\item [$\G, \BG$] Set of all generators and array of sets of generators at bus $i$, indexed by $k$.
\item [$\BL$] Array of sets of loads at bus $i$, indexed by $l$.
\item [$\BS$] Array of sets of shunts at bus $i$, indexed by $s$.
\item [$\tau$] Set of time steps, indexed by $t$ and $h$. 
\item [$\Xi^{[\cdot]}$] Set of decision variables.
\end{ldescription}

\subsection{Parameters}
\begin{ldescription}{$xxxxxxxxx$}
\item [$\CC,\C,\cc$] Generator cost coefficients.
\item [$\dP,\dQ$] Active and reactive power load.
\item [$\gs,\bs$] Bus shunt conductance and susceptance.
\item [$\gl,\gf,\gt$] Branch $\pi$-section conductances.
\item [$\bl,\bf,\bt$] Branch $\pi$-section susceptances.
\item [$\tap,\shift$] Branch tap magnitude and shift angle.
\item [$\Pmin,\Pmax$] Generator minimum and maximum active power production.
\item [$\Qmin,\Qmax$] Generator minimum and maximum reactive power production.
\item [$\Sline$] Branch maximum apparent power.
\item [$\Vmin,\Vmax$] Bus minimum and maximum voltage magnitude.
\item [$\Vtn,\fitn$] Assumed bus voltage magnitude and angle operating points.
\item [$\condv,\condc$] Boolean parameters which indicate whether to use the quadratic form of the voltage and the cosine representations, respectively.
\item [$\conds$] Boolean parameter indicating if the branch power limit is imposed.
\item [$\SoEmax$] Energy storage capacity.
\item [$\qchmax$] Energy storage maximum power.
\item [$\etach,\etadis$] Energy storage (dis)charging efficiency.
\end{ldescription}

\subsection{Variables}
Continuous variables
\begin{ldescription}{$xxxxxxxxx$}
\item [$\Pgen,\Qgen$] Generator active and reactive power production.
\item [$\P,\Q$] Branch active and reactive power flow.
\item [$\dVn,\dfin$] Bus voltage magnitude and angle change.
\item [$\cosf$] Cosine approximation.
\item [$\Va$] Second-order Taylor series voltage magnitude term approximation.
\item [$\SoE$] Energy storage state-of-energy.
\item [$\qbat, \qqbat$] Energy storage active and reactive power.
\item [$\qch, \qdis$] Energy storage (dis)charging active power.
\end{ldescription}

Binary variables
\begin{ldescription}{$xxxxxxxxx$}
\item [$\xch$] Disables simultaneous charging and discharging of energy storage.
\end{ldescription}

\section{Introduction}
\label{sec:intro}

Increasing the integration of renewable energy resources, as well as the electrification trends in the context of the zero-carbon energy future, pushes the power system operation to its technical limits. Operational challenges and increased needs for system flexibility require advances towards new, close to real-time, market products. However, the existing market bidding models do not adequately model the relevant technical aspects, such as voltage magnitudes, node angles and loses, as they rely on simplifications or approximations in the optimal power flow models. Although market designs both in Europe and the US still employ the DC optimal power flow (DC OPF) to perform market clearing, there is a growing interest for using AC optimal power flow (AC OPF), primarily to gain a more accurate insight into the network state and enable a more complete market design that accounts for ancillary services, see e.g. \cite{slp} and \cite{market}. 

Due to the interest of the power industry, there has been a growing interest for development of tractable and accurate AC OPF models. Unlike the linear DC OPF problem, where linearity of the power flow approximation results in good numerical computability even with many binary variables, the non-convexity of the AC OPF problem makes it computationally challenging. The AC OPF models can be grouped into three categories: exact models, relaxations and approximations.

Exact OPF models are based on an exact power flow function, most commonly expressed either using rectangular coordinates \cite{rectangular}, as a nonconvex quadratic optimization problem, or polar coordinates \cite{polar}, as a nonlinear function. There is also the IV (current-voltage) rectangular formulation \cite{iv}, typically used for modeling loads with irregular power-voltage curves, also known as ZIP loads. On the other hand, relaxation models provide an upper bound to the objective function value of the exact models. Thus, if an exact optimization achieves the same objective function value a relaxation model, then this solution is necessarily the global one. Otherwise, global optimality can not be guaranteed.

One of the first relaxation models, which was developed by Jabr \cite{jabr}, is based on the second-order cone relaxation (SOCP), which can be used to relax any nonconvex quadratic formulation. It achieves good results in radial distribution networks due to the absence of the closed loop angle consistency criterion. However, when applied to the transmission network, it results in objective function errors of app. 2\% \cite{pglib}, which is comparable to the total transmission network losses. An extension of the Jabr’s model, developed by Coffrin, and called quadratically constrained relaxation (QC) \cite{QC}, utilizes the McCormick envelopes to tighten the feasibility area based on maximum bus voltage angle differences. The QC model is at least as tight relaxation as the SOCP relaxation \cite{RelaxComp}. Shor’s semidefinite relaxation (SDP) \cite{SDP} is numerically the most demanding out of the commonly used models. It is also at least as tight as the SOCP relaxation \cite{RelaxComp} and usually more accurate than the QC relaxation. The QC model is not considered in this paper since it normally results in similar accuracy as the Jabr's model, but at an increased computation time \cite{RelaxComp}. Shor's SDP model is also not considered, as there are no solvers that can solve a combination of nonlinear and SDP optimizations incurred by the solution techniques presented in the Part II paper.

In approximations, unlike relaxations, the objective function can deviate both positively and negatively from the global optimum. The commonly used DC model, a linear model, also belongs to this category. Linear models in general have difficulties with modeling active power losses since they are in quadratic dependence on the voltage magnitude difference. To prevent negative active power losses, the linear model from \cite{DC_Q} uses penalty factors in the objective function. The model from \cite{LinTaylor} approaches the problem of quadratic losses by introducing nonconvex piecewise linear losses, but at an expense of binary and integer variables. Similar approaches are also published in \cite{R1.4.1} and \cite{R1.4.2}. On the other hand, the linear AC model (LPAC) from \cite{LPAC} approximates the quadratic function with a series of linear inequalities that form a convex space and, thus, do not require binary variables. Except the approximation errors, this model also exhibits relaxations errors due to inequality constraints that replace the intended equality bound. The convex quadratic model (QPAC) was published in 2013 as a still active patent \cite{qpac_patent}. If there are no errors due to deviation from the inequality bounds, this model approximates the power flows well. However, occurrence of these errors depends on the system state and network configuration. 

From the reviewed models, only LPAC can potentially increase its accuracy by using a presolve to approximate the operating point data. Unfortunately, it does not fully use this potential and sometimes it is even more accurate when using the cold start assumptions. This is in detailed discussed in our previous work \cite{cpsota}, demonstrating the benefits of warm start information. In this paper we extend that model by applying it to applicability to the bilevel problems with AC OPF in the lower level. Convex polar second-order Taylor approximation of AC power flows (CPSOTA) from \cite{cpsota} achieves high accuracy by incorporating both the voltage magnitude and the angle second-order Taylor components and by replacing some of the quadratic inequality constraints with linear equality constraints to avoid constraint relaxation errors due to convexification. The replacement is determined by the presolve also developed in \cite{cpsota}. CPSOTA acts as a local AC OPF approximation at a reference operating point which can be well estimated in a bilevel optimization by simply assuming no market influence of the strategic market player. Good numerical tractability of the proposed method is, however, a result of the smoothing techniques presented in Part II, Section II.G.

%Different AC OPF formulation have been applied for solving current challenges of the variable nature of renewable energy sources by including aspects of operational uncertainties \cite{FeasGuarantee} and security of maintaining the supply under contingency criterion \cite{florin}. The authors in \cite{3level} develop a three level problem using QC model of the AC OPF and solve the inner problem using the primal-dual hybrid gradient and analyze three different classes of decomposition cuts. However, all the mentioned paper focused on operational aspects. Very few papers address the aspect of bilevel modelling in the context of market bidding, where there are two conflicting objective function of different system entities.

\begin{figure}[tb]
  \centering
  \includegraphics[scale=0.6,trim={0cm 0cm 0cm 0cm},clip]{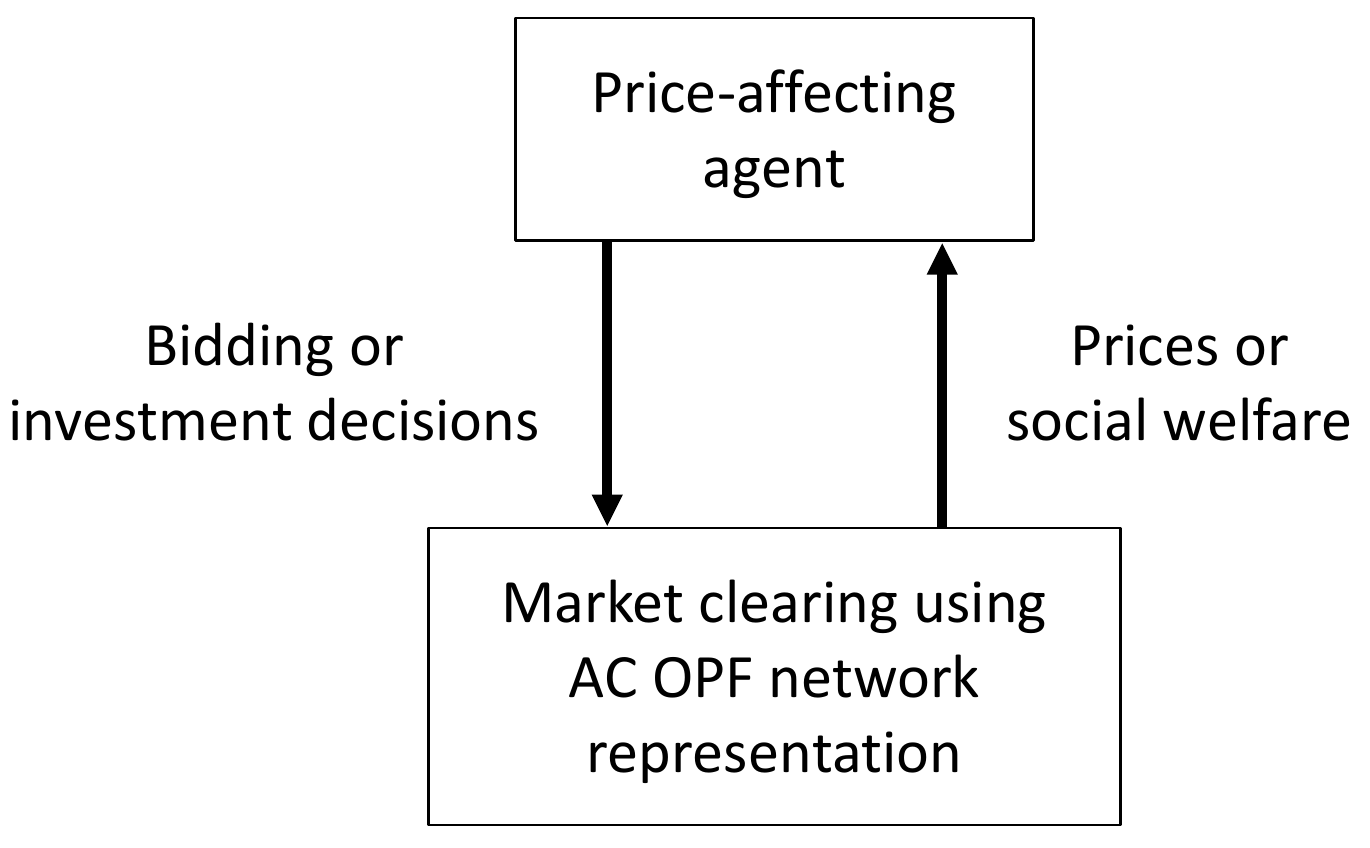}
  \caption{Visualization of the upper-level and lower-level interaction in bilevel problems with market clearing based on AC OPF in the lower level.}
  \label{slika_uvod}
\end{figure}

Convex optimal power flow formulations are a foundation for strategic bidding models relying on bilevel optimization. Generally, in such models the upper-level problem determines the optimal bidding strategy of a strategic player, i.e. the player whose bidding decisions can affect market prices. This player's bidding prices or quantities affect the power flows and market prices. The lower level is used to simulate the market outcome pertaining to the strategic player's bidding actions, as visualized in Figure \ref{slika_uvod}. Examples of such bilevel models are bidding of large consumers \cite{consumer}, generators \cite{generators}, energy storage \cite{ES_Operation} and aggregators \cite{aggregators}. The considered markets may range from the day-ahead market, such as in \cite{consumer}--\cite{aggregators}, to a number of markets, such as in \cite{Stratigic_ES}, where the strategic agent takes part in the day-ahead as well as in the reserve/balancing market. Bilevel models can also be used to determine an optimal investment by considering a number of representative market clearing days within a year. As an example, in \cite{invest_generator} the authors seek an optimal generation investment considering the expected market-clearing conditions.

Strategic agents are not the only ones that use bilevel problems with market clearing in the lower level. The system operators could be interested in increasing the social welfare by investing in new transmission lines \cite{bilevel_TEP}. As an example, in \cite{R1.3.1} and \cite{R1.3.2} the authors consider transmission expansion planning where the SOCP relaxation of the AC OPF is formulated in the lower level. Both papers compare the results only to the option where the lower-level model is replaced with the DC OPF approximation and do not discuss the aspects of achieving accuracy of the exact formulations. More complex models include three levels, one considering the independent agent investments, another considering the system operator investments, and the final one to simulate market outcomes. For example, \cite{ES_invest} formulates a trilevel model where the upper-level problem optimizes the system operator's transmission line and energy storage investments, the middle-level problem determines the merchant energy storage investment decisions, and the lower-level problem simulates a market clearing process for representative days using DC OPF. There are multiple ways of solving trilevel problems, but the first step is always to merge the middle- and lower-level problem into a single-level equivalent.

For conciseness, in this paper we consider a bilevel structure, however, the described procedure can be employed to trilevel problems as well. Also for brevity reasons, we choose a simple energy storage bidding model in the day-ahead market as the upper-level problem and focus on the AC OPF in the lower-level problem and on effectively converting and solving the initial problem.

For a strategic player (or a system operator) to solve a bilevel problem, it needs to first convert it into a single-level equivalent. However, the conversion techniques do not hold for nonconvex models. An excellent review of bilevel optimization approaches can be found in \cite{tutorial} and \cite{tutorial2}, where different techniques, later applied for solving the problem described in this paper, are explained in detail. Traditionally, the reduction techniques of transforming a bilevel problem into a single-level one are based on the primal-dual theory and the Karush–Kuhn–Tucker optimality conditions \cite{decomp}. Both techniques follow the idea of replacing the lower-level problem with a set of equations and inequalities that have the same solution as the original lower-level problem. This set of equations is then added as a set of constraints to the upper-level problem, finally forming a single-level problem equivalent to the initial bilevel problem. 

The main obstacle in the AC OPF formulations for strategic bidding are complementarity condition constraints, which are difficult for any interior point solver since they do not satisfy the Mangasarian-Fromovitz constraint qualification, meaning there is no strictly feasible point, making the nonlinear programming numerically unstable \cite{qualification}. This aspect is one of the reasons why, to the best of the authors' knowledge, all existing bilevel market models rely either on the DC approximation \cite{DCbilevel} or linearized AC models \cite{AClin}. This is because in these cases the complementary conditions can be reshaped into a mixed-integer linear form and solved using the branch-and-bound method. We found only two publications that either explicitly address or can be generalized to bilevel formulations and include exact, relaxed or approximate AC OPF formulations in the lower-level problem. The authors of \cite{contingency} propose a bilevel problem of the worst contingency under attack, where they address the nonlinearities of the AC model as convex SOCP and SDP relaxations. However, this single-level reduction approach is applicable only because the interaction variable between the upper level and the lower level is binary, thus allowing for an exact reformulation of the resulting bilinear terms using the McCormick envelopes. In the second one \cite{slp}, the authors propose a successive linear algorithm based on the dual form of a linear Taylor expansion of the IV-ACOPF model to solve the AC OPF. Their approach can be expanded to bilevel modeling with AC OPF-constrained market clearing in the lower level, however, their algorithm on the presented 14\_ieee network does not converge. After six iterations and a DC OPF-based starting point, the LMP oscillations are at 0.20\%. If the upper level is added to the model, its interaction with the lower level could potentially further destabilise convergency. It also uses generally nontrivial constraint violation penalties and constrains maximum deviations of the Taylor delta variables.

%Recent mathematical progress provides numerically more tractable and stable approach of formulating complementary condition constraints by using smoothing functions \cite{sm}. In this paper we avoid model linearization and use convex quadratically constrained quadratic formulation instead.

%The quality of the convergence and the results are benchmarked against the DC OPF problem, however there is no further discussion on the exactness, accuracy or computational tractability when applied to different and real networks. Also, a comparison to other methods is missing.

\begin{figure}[tb]
  \centering
  \includegraphics[scale=0.4,trim={0cm 0cm 0cm 0cm},clip]{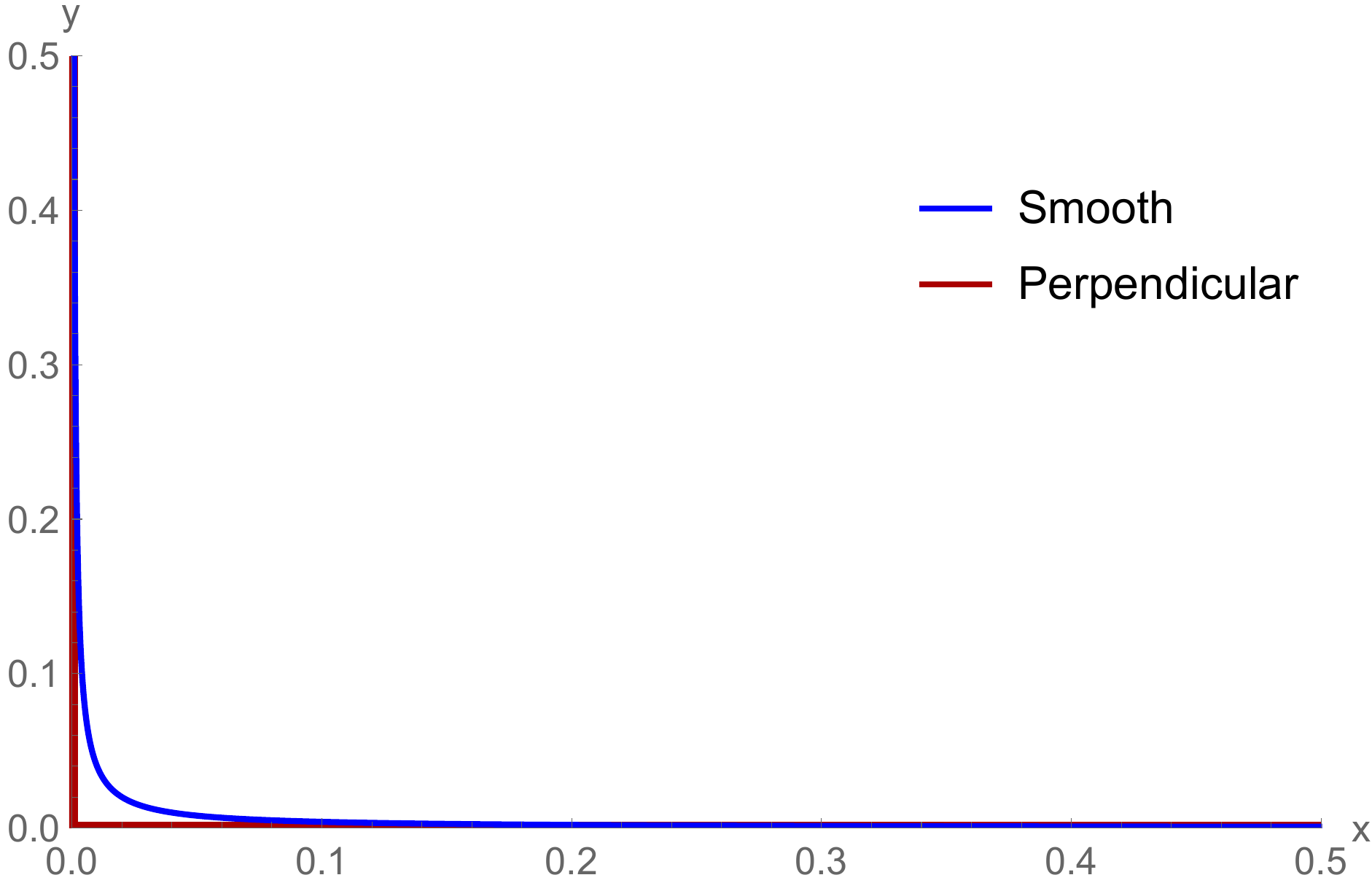}
  \caption{Smoothing example of the perpendicular relation.}
  \label{fig:smooth}
\end{figure}

Besides the techniques listed above, there are some interesting mathematical tools whose applicability has not yet been explored in the power systems community. In paper \cite{sm}, the authors examine the Lipschitzian and differential properties of smoothing functions and their application to optimization with SOC complementarity constraints. Application of the smoothing techniques still results in nonlinear expressions, however, much easier to solve since they have all derivations in every point of the function. An example of smoothing a linear complementary condition is shown in Figure \ref{fig:smooth}. The perpendicular lines represent the following three constraints $x\ge 0$, $y\ge 0$ and $x\! \cdot \! y = 0$, while the corresponding smooth curve is represented by only a single smooth constraint $x+y-\sqrt{x^2+y^2+2\eps^2}=0$. Despite the promising theoretical foundations of the reformulated smooth constraints to achieve good numerical tractability with interior point based solvers, the technique has not yet been practically used or demonstrated. In this paper we employ the novel smoothing techniques, first proposed and developed in \cite{sm}, to tractably solve bilevel problems with AC OPF in the lower level. This allows us to avoid model linearization and use any AC OPF formulation of continuous SOCP class or simpler. Smoothing technique implementation details are explained in Section II.G of Part II of this work.

Based on the above, the paper brings the following original contribution:
\begin{itemize}
    \item For the first time we develop and present a mathematical formulation of a bilevel problem based on the smoothing techniques, where a strategic player's profit maximization is in the upper level, while the AC OPF problem is the lower level. For demonstration purposes, we choose energy storage (ES) as the strategic player making the bidding decisions in the upper-level problem. However any other strategic player, e.g. generator, demand response, aggregator of different flexibility sources, or a system operator can be plugged into the upper level. Furthermore, the model can be easily expanded into an investment model that considers multiple representative days.
    \item The bilevel problem is reduced to a single-level problem using the known techniques, i.e. primal-dual counterpart, strong duality theorem, McCormick envelopes, complementary slackness, penalty factor and interaction discretization. For all these techniques the computational tractability, effort and accuracy are analyzed and issues are detected and elaborated.
    \item All the developed models and codes are made available as open access to the scientific community at \cite{source}.
\end{itemize}

%The exactness, accuracy and computational time of the proposed method is demonstrated on a small three-bus network and a large case study.

Rest of the paper is structured as follows. Section \ref{sec:math} mathematically states the proposed model. It is divided into two parts: Subsection \ref{sub:initial} presents the initial model, which is reformulated into its SOC form in Subsection \ref{sub:reformulation}. Section \ref{sec:presolve} introduces the algorithm and presolve to obtain the model's prerequisites and verifies the accuracy of the obtained solution. Solution techniques and the case studies are presented in the companion paper.%Section IV presents solution techniques we employ to solve the bilevel problem. The case study section V consists of three main parts: the description and set-up subsection V-A; and the two case studies. The first case study V-B demonstrates the model's accuracy, while the second V-C presents an in-depth solution techniques analysis. The final section VI provides conclusive remarks.

\section{Mathematical Model}\label{sec:math}

The presented bilevel optimization model consists of a simple ES active and reactive power bidding model in the upper level and the AC OPF in the lower level. Mathematical focus is on demonstrating an accurate and computationally tractable bilevel AC OPF solution.

\subsection{Initial Model}
\label{sub:initial}

\subsubsection{Upper Level}\hfill

The upper level considers a large ES unit located at bus $i\in \BB$. The objective function \eqref{eq:UL.1} maximizes its profit by performing the day-ahead energy and reactive power market arbitrage while its impact on the prices is acknowledged by the dual variables $\lambdaJedan$ for active power and $\lambdaDva$ for reactive. Constraint \eqref{eq:UL.2} models the storage (dis)charging process considering its efficiency, while \eqref{eq:UL.3}--\eqref{eq:UL.5} limit the maximum ES capacity, charging and discharging rates. Constraint \eqref{eq:UL.6} combines the charged and discharged energy into a cumulative quantity $\qbat$. Binary variable $\xch$ disables simultaneous charging and discharging which could otherwise occur in periods with negative prices. However, in many cases in the case study $\xch$ is dropped-out to evaluate the performance of the solution techniques on a fully continuous model. When $\xch$ is dropped out, constraints \eqref{eq:UL.4} and \eqref{eq:UL.5} are conceptually not needed due to constraint \eqref{eq:UL.7} which limits ES the apparent power.

\begin{equation}\tag{1.1}
\underset{\Xiul}{\mathrm{Max}} \sum_{t,i\in \BB} (\qbat \cdot \lambdaJedan + \qqbat \cdot \lambdaDva)
\label{eq:UL.1}
\end{equation}

\begin{equation}\tag{1.2}
\SoE = \SoEm + \qch\cdot \etach - \qdis/\etadis, \quad \forall{t}
\label{eq:UL.2}
\end{equation}

\begin{equation}\tag{1.3}
0 \le \SoE \le \SoEmax, \quad \forall{t}
\label{eq:UL.3}
\end{equation}

\begin{equation}\tag{1.4}
0 \le \qch \le \qchmax \cdot \xch, \quad \forall{t}
\label{eq:UL.4}
\end{equation}

\begin{equation}\tag{1.5}
0 \le \qdis \le \qdismax \cdot (1-\xch), \quad \forall{t}
\label{eq:UL.5}
\end{equation}

\begin{equation}\tag{1.6}
\qbat = \qch - \qdis, \quad \forall{t}
\label{eq:UL.6}
\end{equation}

\begin{equation}\tag{1.7}
(\qbat)^2+(\qqbat)^2 \le (\qchmax)^2, \quad \forall{t}
\label{eq:UL.7}
\end{equation}

The upper-level set of variables is $\Xiul=\{\qbat,\qch,\qdis,\qqbat,\SoE,\xch\}$. $\xch$ is an optional binary variable that can be neglected in case of nonnegative market prices.

\subsubsection{Lower Level}\hfill

%Loads, generators and storages take part in electricity markets to purchase or sell energy. 

Transmission-constrained electricity markets perform market clearing by running an OPF optimization. Here we model the network using the CPSOTA \cite{cpsota}. The CPSOTA model is chosen due to convexity which is needed to satisfy the single-level reduction regularity conditions and due to superb warm-start transmission system OPF accuracy as demonstrated in \cite{cpsota} as well as in the case study of the second part of this paper. The CPSOTA model is designed to exploit a meshed network structure to achieve accuracy and should not be used with radial networks. Both the loads' and ES bids are considered inelastic, meaning they will always be cleared. This assumption holds if the loads (including the ES when buying) bid very high prices and the ES bids very low price when selling energy. Thus, both the loads and the ES (dis)charged energy $\qbat$ are modeled as parameters in the lower level. This simplification is introduced to shorten the upper-level objective function convexification (4.1) and its follow-up versions appearing in the Part II paper.

%and not due to inability to model traditional bids.

The lower level minimizes the quadratic production costs in its objective function \eqref{eq:LL.1}. It is assumed that generators bid reactive power at zero prices. Bus power balance is imposed in \eqref{eq:LL.2} and \eqref{eq:LL.3}. The upper-level variables $\qbat$ and $\qqbat$ appear in constraints \eqref{eq:LL.2} and \eqref{eq:LL.3} only for the bus at which the ES is located, i.e. when condition $:\!i\!\in\! \BB$ under the variable is true. Our constraint writing style also assumes that summation indices are first fixed by the outer \textit{for all} statement and then the summation rolls over to the remaining unfixed indices. This way $\sum_{(e,i,j)\in \A}\P$, $\forall{t,i}$ sums the active powers from all branches originating from bus $i$. Constraints \eqref{eq:LL.4}--\eqref{eq:LL.7} are active and reactive power flow equations. Since these are based on the Taylor expansion, the computed voltage magnitude is equal to the assumed operating point value $\Vtn$ plus deviation $\dVn$. Similarly, the computed voltage angle is $\fitn+\dfin$. To shorten the expressions, parameters $\cospsinij$ and $\cosmsinij$ are defined over the other basic parameters, i.e. the ones present in the nomenclature, as introduced in the Appendix \eqref{eq:P.1.1}--\eqref{eq:P.2.2}. Constraints \eqref{eq:LL.8.1} and \eqref{eq:LL.9.1} are second-order Taylor voltage magnitude and angle approximations, respectively, whose variables $\Va$ and $\cosf$ also appear in the power flow equations. To achieve convexity, these constraints are inequalities as no quadratic equality is convex. At the solution point, it is intended that these constraints are binding so that no errors occur due to a swap of the equality sign with inequality. To achieve this, there is a presolve step in the optimization, as described in Section \ref{sec:presolve}, which decides on swapping the quadratic inequality constraints that are likely to diverge from the inequality boundary with simple linear equality constraints \eqref{eq:LL.8.2} and \eqref{eq:LL.9.2}, thus avoiding gross errors. Applying the linear equality constraint variants disregards the second-order part of the Taylor. Even the zero-order Taylor expansion is exact at the expansion point, so disregarding a few second-order terms on per-branch or per-bus pair basis does not worsen the solution significantly in the vicinity of the expansion point. Constraints \eqref{eq:LL.10} and \eqref{eq:LL.11} limit the generators' production capabilities. Constraint \eqref{eq:LL.12} limits the maximum branch apparent power. Since \eqref{eq:LL.12} is a major source of computationally demanding quadratic equations, the presolve reduces their number for the main solve using a preset threshold at the initial operating point. More on the multi-step presolve procedure which finds an approximate operating point $\Vtn$, $\fitn$ and decides on the use of quadratic constraints, i.e. determines the values of the Boolean parameters $\condv$, $\condc$ and $\conds$, can be found in Section \ref{sec:presolve}. Finally, constraint \eqref{eq:LL.13} sets the reference bus angle to zero and constraint \eqref{eq:LL.14} sets the voltage magnitude bounds.

\begin{equation}\tag{2.1}
\underset{\Xill}{\mathrm{Min}} \; \Omegap := \sum_{t,k}(\CC\cdot (\Pgen)^2 + \C \cdot \Pgen + \cc)
\label{eq:LL.1}
\end{equation}

\begin{equation}\tag{2.2}
\begin{split}
&\sum_{k \in \BG}\Pgen - \sum_{l\in \BL}\dP -\hspace{-13pt} \sum_{(e,i,j)\in \A}\hspace{-13pt}\P - \underset{\hspace{-7pt}:i\in\BB}{\qbat}\\[4pt]
&- ((\Vtn)^2\!+\!2\!\cdot\!  \Vtn\! \cdot \! \dVn)\cdot\!\!\! \sum_{s\in \BS}\!\gs \!=\! 0, \!\!\quad \forall{t,i} \!\!\quad :\lambdaJedan
\label{eq:LL.2}
\end{split}
\end{equation}

\begin{equation}\tag{2.3}
\begin{split}
&\sum_{k \in \BG}\Qgen - \sum_{l\in \BL}\dQ -\hspace{-13pt} \sum_{(e,i,j)\in \A}\hspace{-13pt}\Q - \underset{\hspace{-7pt}:i\in\BB}{\qqbat}\\[4pt]
&+ ((\Vtn)^2\!+\!2\!\cdot\!  \Vtn\! \cdot\! \dVn)\cdot\!\!\! \sum_{s\in \BS}\!\bs\! =\! 0, \!\!\quad \forall{t,i} \!\!\quad :\lambdaDva 
\label{eq:LL.3}
\end{split}
\end{equation}

\begin{equation}\tag{2.4}
\begin{split}
&\P= ((\Vtn)^2\!+\!2\!\cdot\!\Vtn\!\cdot\!\dVn)\!\cdot\! (\gl\!+\!\gf)/\tap^2 \!+\! \Va/2\\[4pt]
&\!-\!\cospsinij \!\cdot\! (\Vtn\!\cdot\! \Vtm \!\cdot\! \cosf \!+\! \dVn\! \cdot\! \Vtm  \!+\! \dVm \!\cdot \!\Vtn)/\tap\\[4pt]
&\!-\!\cosmsinij \!\cdot \!\!\Vtn\!\!\cdot\! \Vtm \!\!\cdot \! (\dfin \!-\! \dfim) /\tap,\\[4pt]
&\hspace{100pt} \forall{t,(e,i,j)\in \AF} \quad :\lambdaTri\\[-0.9cm]
\label{eq:LL.4}
\end{split}
\end{equation}

\begin{equation}\tag{2.5}
\begin{split}
&\P= ((\Vtn)^2\!+\!2\!\cdot\!\Vtn\!\cdot\!\dVn)\!\cdot\! (\gl\!+\!\gt) \!+\! \Va/2\\[4pt]
&\!-\!\cospsinij \!\cdot\! (\Vtn\!\cdot\! \Vtm \!\cdot\! \cost \!+\! \dVn\! \cdot\! \Vtm  \!+\! \dVm \!\cdot \!\Vtn)/\tap\\[4pt]
&\!-\!\cosmsinij \!\cdot \!\!\Vtn\!\!\cdot\! \Vtm \!\!\cdot \! (\dfin \!-\! \dfim) /\tap,\\[4pt]
&\hspace{100pt} \forall{t,(e,i,j)\in \AT} \quad :\lambdaCetiri\\[-0.9cm]
\label{eq:LL.5}
\end{split}
\end{equation}

\begin{equation}\tag{2.6}
\begin{split}
&\Q= -((\Vtn)^2\!+\!2\!\cdot\!\Vtn\!\cdot\!\dVn)\!\cdot\! (\bl\!+\!\bf)/\tap^2\\[4pt]
&\!+\!\cosmsinij \!\cdot\!(\Vtn\!\cdot\! \Vtm\! \cdot\! \cosf \!+\! \dVn \!\cdot \!\Vtm  \!+\! \dVm \!\cdot\! \Vtn)/\tap\\[4pt]
&\!-\!\cospsinij \!\cdot\! \Vtn\! \cdot\! \Vtm \!\!\cdot \!(\dfin \!-\! \dfim) /\tap,\\[4pt]
&\hspace{100pt} \forall{t,(e,i,j)\in \AF} \quad :\lambdaPet\\[-0.90cm]
\label{eq:LL.6}
\end{split}
\end{equation}

\begin{equation}\tag{2.7}
\begin{split}
&\Q= -((\Vtn)^2\!+\!2\!\cdot\!\Vtn\!\cdot\!\dVn)\!\cdot\! (\bl\!+\!\bt)\\[4pt]
&\!+\!\cosmsinij \!\cdot\! (\Vtn\!\cdot\! \Vtm\! \cdot \! \cost \!+\! \dVn \!\cdot\! \Vtm  \!+\! \dVm \!\cdot\! \Vtn)/\tap\\[4pt]
&\!-\!\cospsinij \!\cdot\! \Vtn\!\cdot \!\Vtm \!\!\cdot\! (\dfin \!-\! \dfim) /\tap,\\[4pt]
&\hspace{100pt} \forall{t,(e,i,j)\in \AT} \quad :\lambdaSest\\[-0.90cm]
\label{eq:LL.7}
\end{split}
\end{equation}

\begin{equation}\tag{2.8.1}
\begin{split}
&\Va \!\!\ge\!\! (\gl\!\!+\!\hspace{-0.7pt}\gf\hspace{-0.5pt}) \!\hspace{-1pt} \cdot \!\hspace{-1pt}(\hspace{-0.5pt} \dVn \hspace{-0.5pt})^2\! \hspace{-0.5pt} /\hspace{-0.5pt}\tap^2\!\hspace{-1.3pt}-\!\!2\!\cdot\! \gl\!\!\cdot\! \mathrm{cos}(\fitn\!\!-\!\fitm\!\!-\!\shift) \!\cdot \!\! \dVn \!\!\cdot\!\! \dVm/\tap \\[4pt]
&+\!(\gl\!+\!\gt)\!\cdot\! (\dVm)^2,  \quad \forall{t,(e,i,j)\!\in \! \AF} :\condv\\[-0.95cm]
\label{eq:LL.8.1}
\end{split}
\end{equation}

\begin{equation}\tag{2.8.2}
\Va = 0, \quad \forall{t,\!(e,i,j)\!\in\! \AF} : \lnot \condv \!\!\quad : \lambdaDvadesetjedan
\label{eq:LL.8.2}
\end{equation}

\begin{equation}\tag{2.9.1}
\!\cosf \le 1 - (\dfin-\dfim)^2/2, \quad \forall{t,(i,j)\in \BP} : \condc
\label{eq:LL.9.1}
\end{equation}

\begin{equation}\tag{2.9.2}
\cosf = 1, \quad \forall{t,(i,j)\in \BP} : \lnot \condc \!\!\quad : \lambdaDvadesetdva
\label{eq:LL.9.2}
\end{equation}

\begin{equation}\tag{2.10}
\Pmin\! \le \Pgen \le \Pmax , \quad \forall{t,k} \quad : \muJedanDn, \muJedanUp
\label{eq:LL.10}
\end{equation}

\begin{equation}\tag{2.11}
\Qmin\! \le \Qgen \le \Qmax , \quad \forall{t,k} \quad : \muDvaDn, \muDvaUp
\label{eq:LL.11}
\end{equation}

\begin{equation}\tag{2.12}
\begin{split}
\P^2\!+\!\Q^2 \! \le &\Sline^2, \!\!\quad \forall{t,\!(e,i,j)\!\in\! \A } :\conds,\\
&:\lambdaJedanaest,\lambdaDvanaest, \lambdaTrinaest
\label{eq:LL.12}
\end{split}
\end{equation}

\begin{equation}\tag{2.13}
\fitn + \dfin = 0, \quad \forall{t,i \in \RB} \quad : \lambdaSedam
\label{eq:LL.13}
\end{equation}

\begin{equation}\tag{2.14}
\Vmin \le \Vtn+\dVn \le \Vmax, \quad \forall{t,i} \quad : \muTriDn,\muTriUp
\label{eq:LL.14}
\end{equation}

Lower level set of primal variables: $\Xill=\{\dfin,\dVn,\P,$ $\Q,\Pgen,\Qgen\,\cosf,\Va\}$.

\subsection{SOC Constraint Reformulation}
\label{sub:reformulation}

To derive the dual problem needed for a single-level reduction, quadratic primal constraints are first converted into their equivalent SOC form. Otherwise, the direct conversion would result in a general nonlinear formulation provided in \cite{ejor_qcqp}, as opposed to our SOCP dual. Thus, this conversion is essential for solution techniques reliant on the SOCP form, i.e. primal-dual counterpart, McCormick envelopes, interaction discretization and smoothing techniques from the Part II paper. A SOC constraint conversion is nontrivial because there exist infinitely many SOC formulations of the same convex quadratic constraint. On the other hand, there also exists a formula with a unique positive semidefinite solution for conversion of the convex quadratic constraints to the SOC form \cite{conversion}. However, we decided to manually choose our own SOC form since the general formula can result in a form with more than minimum possible number of cone variables and extensively complicated coefficients.

The resulting basic cone constraints are \eqref{eq:SOC.1.1} and \eqref{eq:SOC.2.1}. Together with the substitution defining \eqref{eq:SOC.1.2}--\eqref{eq:SOC.1.5} and \eqref{eq:SOC.2.2}--\eqref{eq:SOC.2.4}, they are equivalent to the initial quadratic voltage magnitude \eqref{eq:LL.8.1} and angle \eqref{eq:LL.9.1} constraints, respectively. To shorten the expressions, coefficients, i.e. parameters, $\vpJedan$, $\vpDva$ and $\vpTri$ are defined in the Appendix \eqref{eq:P.3}--\eqref{eq:P.5}. Also, it is useful to note that the basic cone constraints share the same dual variable with their right-hand-side substitution constraint.

The remaining quadratic parts of the model are the branch apparent power limit constraint \eqref{eq:LL.12}, which is already in the SOC form, and the objective function \eqref{eq:LL.1}. To derive the dual, we recognize three ways of dealing with a quadratic objective function: \begin{itemize}
    \item apply the QP duality theory;
    \item transform it into a single large (multidimensional) SOC constraint;
    \item transform it into multiple three dimensional SOC constraints.
\end{itemize}

In this work we apply the QP duality theory to the objective function, while the rest of the model is converted into the dual using the SOCP procedure. By the QP duality theory, a dual is derived by writing the Karush–Kuhn–Tucker (KKT) conditions and then eliminating the remaining primal variables, which remain after derivations due to the squares in the primal objective function, by substituting them from the KKTs into the Lagrange function as described in lecture \cite{qcdual}. The SOC constraints are converted into the dual by using the mathematical theorem that states that the basic cones are self-dual \cite{selfdual}, i.e. for every primal basic SOC constraint there is a dual basic SOC constraint (see primal-dual constraint pair (4.10) and (4.11) from Part II paper). Other constraints are linear and their KKT conditions can be used to derive the dual. This way the dual model is of the same optimization class as the primal, i.e. the objective function is convex quadratic and the constraints are SOC. The write-out of the dual is available in the Appendix. The two other objective function transformation approaches are not preferred over the QP procedure since they enlarge both the primal and dual models and, since the SOC constraints are inequalities, introduce additional complementary slackness conditions.

\begin{equation}\tag{3.1.1}
\vvJedan^2 \hspace{-1pt} \!+\! \hspace{-0.5pt}  \vvDva^2 \hspace{-1pt} \!+\! \hspace{-0.5pt} \vvTri^2 \hspace{-1pt}\!\le\!\hspace{-1pt} \vvCetiri^2, \!\!\quad\forall{t,\!(e,i,j)\!\in \! \AF} \!:\!\condv
\label{eq:SOC.1.1}
\end{equation}

\begin{equation}\tag{3.1.2}
\vvJedan \!=\! (1 - \Va)/2, \!\quad \forall{t,\!(e,i,j)\!\in \! \AF} : \condv \!\quad : \lambdaSedamnaest
\label{eq:SOC.1.2}
\end{equation}

\begin{equation}\tag{3.1.3}
\begin{split}
\hspace{-0.25cm}\vvDva \!=\!\vpJedan & \cdot \dVn - \vpDva \cdot \dVm,\\
&\forall{t,(e,i,j)\!\in \! \AF} :\condv \quad : \lambdaOsamnaest
\label{eq:SOC.1.3}
\end{split}
\end{equation}

\begin{equation}\tag{3.1.4}
\vvTri \!=\!\vpTri \cdot \dVn, \quad \forall{t,(e,i,j)\!\in \! \AF} :\condv \quad : \lambdaDevetnaest
\label{eq:SOC.1.4}
\end{equation}

\begin{equation}\tag{3.1.5}
\vvCetiri \!=\!  (1 + \Va)/2,\quad \forall{t,(e,i,j)\!\in \! \AF} : \condv \quad : \lambdaDvadeset
\label{eq:SOC.1.5}
\end{equation}

\begin{equation}\tag{3.2.1}
\cvJedan^2 \!+\! \cvDva^2 \!\le\! \cvTri^2, \quad \forall{t,(i,j)\!\in \! \BP} :\! \condc
\label{eq:SOC.2.1}
\end{equation}

\begin{equation}\tag{3.2.2}
\!\cvJedan \!\!=\!\! (\dfin \!- \dfim)\!/\!\sqrt{2}, \!\!\!\! \quad \forall{t,\!(i,j)\!\in \! \BP} :\! \condc \!\!\!\!\! \quad :\! \lambdaCetrnaest\!\!
\label{eq:SOC.2.2}
\end{equation}

\begin{equation}\tag{3.2.3}
\!\!\cvDva \!\!=\! \cosf \!-\!3/4, \!\!\! \quad \forall{t,(i,j)\!\in \! \BP} \!:\! \condc \!\!\!\! \quad :\! \lambdaPetnaest
\label{eq:SOC.2.3}
\end{equation}

\begin{equation}\tag{3.2.4}
\!\!\cvTri \!\!=\! \!-\cosf + 5/4, \!\!\!\!\! \quad \forall{t,(i,j)\!\in \! \BP} \!:\! \condc \!\!\!\!\!\! \quad :\! \lambdaSesnaest
\label{eq:SOC.2.4}
\end{equation}

The reformulated lower-level set of variables is $\Xir=\Xill \cup \{\vvCetiri,\vvJedan,\vvDva,\vvTri,\cvTri,\cvJedan,\cvDva\}$.

\section{Algorithm and Presolve}
\label{sec:presolve}

The presented bilevel model requires as an input both the numerical and the Boolean parameters that need to be determined beforehand. This section introduces the algorithm to obtain prerequisites and verify the accuracy of the obtained solution.

The lower-level problem consists of an AC OPF model based on a Taylor expansion and thus requires an approximate operating point for voltage magnitude $\Vtn$ and angle $\fitn$ as inputs. The first step in the Algorithm is to compute it using the exact polar model. The computation is based on the assumption that the strategic player in the upper-level problem is passive, i.e. the energy storage is neither charging nor discharging. Thus, the optimization is a typical, single-level AC OPF. To reduce the computational burden of the main bilevel optimization, the first step determines on which lines in the forthcoming steps the power limits will not be imposed, controlled with Boolean parameter $\conds$, using a preset threshold. In case the final solution overloads some of the lines, the threshold can be changed.

The second step is the presolve, aimed to drastically improve the AC OPF approximation model accuracy. The main potential inaccuracy source is the relaxation of the equality sign in constraints \eqref{eq:LL.8.1} and \eqref{eq:LL.9.1} into the inequality sign to achieve convexity. To prevent relaxation errors in the forthcoming steps, this step marks all the constraints that would at the operating point deviate from the inequality boundary and replaces them with their linear equality variants \eqref{eq:LL.8.2} and \eqref{eq:LL.9.2} controlled with Boolean parameters $\condv$ and $\condc$. This step itself is nonconvex. It reruns the operating point using CPSOTA, but uses exclusively \eqref{eq:LL.8.1} and \eqref{eq:LL.9.1} quadratic constraints as equalities, thus not susceptible to relaxation errors and not requiring any information about the $\condv$ and $\condc$ parameters which it determines for the forthcoming steps. The selection of constraints for the forthcoming steps is based on their marginal value, computed by default by many solvers, e.g. IPOPT and Knitro. A constraint marginal is a sensitivity of the primal objective function on adding a small positive constant to the right-hand side of the constraint and thus its sign indicates whether an equality constraint would be binding if it were relaxed into inequality. In case of the voltage constraint \eqref{eq:LL.8.1}, it would be binding if the marginal is positive and thus $\condv$ is true. The cosine constraint \eqref{eq:LL.9.1} it would be binding if the marginal is negative and thus $\condc$ is true. This step has the same solution as the first one and all delta variables $\dVn$ and $\dfin$ are zero. The described presolve working principle is further explained in \cite{cpsota}. It minimizes \eqref{eq:LL.1} subject to \eqref{eq:LL.2}--\eqref{eq:LL.14} with \eqref{eq:LL.8.1} and \eqref{eq:LL.9.1} as equalities, without \eqref{eq:LL.8.2} and \eqref{eq:LL.9.2}, with respect to the variables set $\Xill$.

The third and fourth steps are convex and simply resolve the primal and dual problems at the determined operating point ($\Vtn$ and $\fitn$) for the selected constraints ($\condv$, $\condc$ and $\conds$) to supply the variables with their warm start values for the bilevel solve. Warm start values initialize the interior-point-based solvers, e.g. initializing Jacobian and Hessian matrices, enhancing the numerical tractability. The strategic player is still considered passive and, at the solution, the objective function values are the same as in the previous steps. The third step solves the SOCP version of the lower-level problem and the fourth step solves the SOCP dual. Specifically, third step minimizes \eqref{eq:LL.1} subject to \eqref{eq:LL.2}--\eqref{eq:LL.14}, excluding \eqref{eq:LL.8.1} and \eqref{eq:LL.9.1} in favor of \eqref{eq:SOC.1.1}--\eqref{eq:SOC.2.4} with respect to the variables set $\Xir$ and the forth step maximizes \eqref{eq:D.1} subject to \eqref{eq:D.2}--\eqref{eq:D.12} with respect to the variables set $\Xidu$.

The sixth step, which comes after the bilevel solve step five, verifies the solution accuracy. It solves the exact polar AC OPF with fixed (dis)charging decisions to the bilevel solve. It determines the actual system expenses and the upper-level profit for decisions from the bilevel problem. For computing the upper-level profit, nodal prices are obtained from the nodal power balance constraint marginal. The described procedure is itemized in Algorithm \ref{alg:bilevel}.

\begin{figure*}[tb]
  \centering
  \includegraphics[scale=0.55,trim={0cm 0.3cm 0cm 0cm},clip]{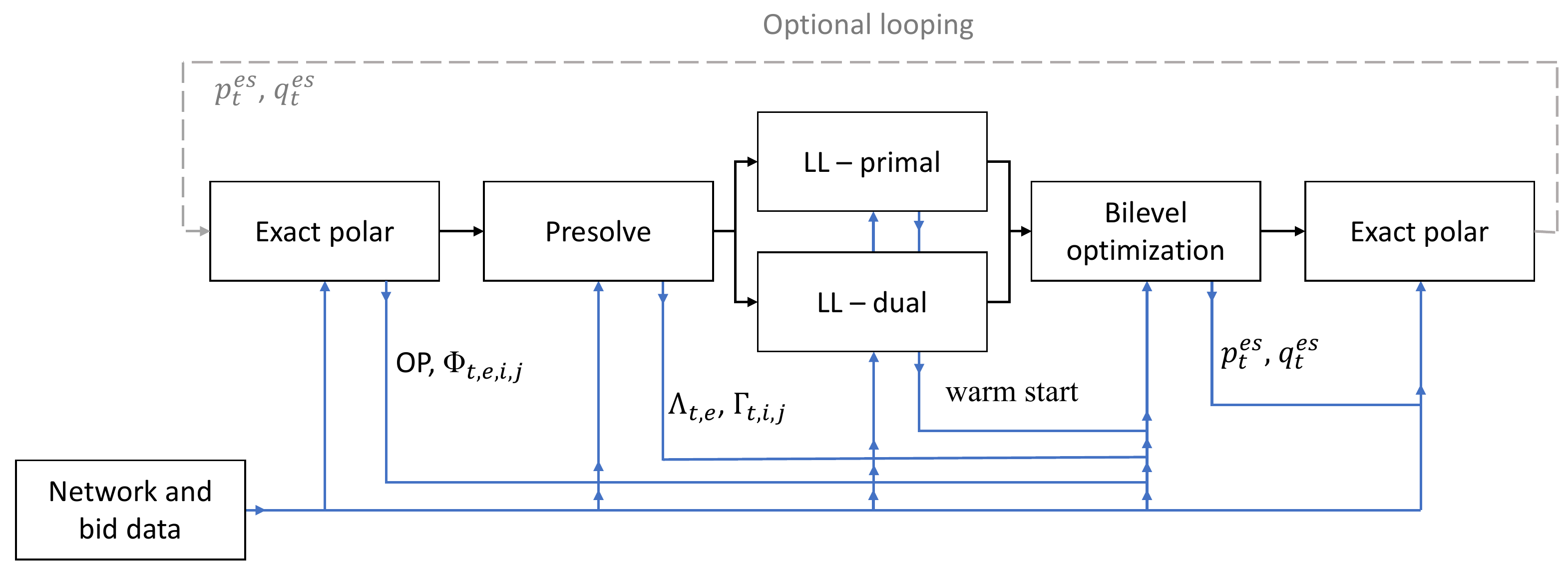}
  \caption{Visualization of the Algorithm 1 and input data.}
  \label{alg1_s}
\end{figure*}

The proposed Algorithm is conceptually iterable to improve accuracy. Steps 1--6 can be run in a loop where the first step computes a new operating point assuming fixed (dis)charging decisions from the last bilevel solve. When looping, Steps 1 and 6 solve the same problem and can be performed in a single optimization. The described Algorithm is visually presented in Figure \ref{alg1_s}.

\begin{algorithm}[bht]
  \caption{Bilevel optimization}\label{alg:bilevel}
  \begin{algorithmic}[1]
%    \Procedure{Euclid}{$a,b$}\Comment{The g.c.d. of a and b}
      \State Run exact polar model \Comment{NLP; determine OP, $\conds$}
      \State Run presolve \Comment{NC-QCQP; determine $\condv$, $\condc$}
      \State Run LL-primal \Comment{SOCP; warm start solver}
      \State Run LL-dual \Comment{SOCP; warm start solver}
      \State Run bilevel optimization\!\! \Comment{\!Opt. \!class \!varies; \!deter. \!$\qbat$\!,\! $\qqbat$\!}
      \State Run exact polar model \Comment{NLP; verify solution}
%    \EndProcedure
  \end{algorithmic}
\end{algorithm}

\section{Conclusion}

This paper presented the mathematical model of a strategic energy storage acting in the day-ahead market with the market clearing based on the AC OPF in the lower level. It presents the SOC constraint reformulation and proposes an algorithm to accurately solve such complex structure in a time-efficient manner. The solution techniques as well as the case studies are presented in the accompanying paper.

\input{appendix3}

%\subsection{Convexified upper-level objective function}
%\label{sub:OF}

\bibliographystyle{IEEEtran}
%\bibliography{bibliography}
%\vspace{-.3cm}

\begin{comment}

\begin{IEEEbiography}[{\includegraphics[width=1in,height=1.25in,clip,keepaspectratio]{sepetanc_index_grey.jpg}}]{Karlo \v{S}epetanc}
received the M.E.E. degree from the University of Zagreb Faculty of Electrical Engineering and Computing, Croatia, in 2018 and is currently pursuing a Ph.D. at the same university. He is also a researcher at the Innovation Centre Nikola Tesla (ICENT).

His research interests include planing, operation, and economics of power and energy systems.
\end{IEEEbiography}

\begin{IEEEbiography}[{\includegraphics[width=1in,height=1.25in,clip,keepaspectratio]{pandzic_IEEE.jpg}}]{Hrvoje Pand\v{z}i\'{c}}
(M'08--SM'17) received the M.E.E. and Ph.D. degrees from the University of Zagreb Faculty of Electrical Engineering and Computing, Croatia, in 2007 and 2011. From 2012 to 2014, he was a Postdoctoral Researcher with the University of Washington, Seattle, WA, USA.

Currently, he is an Associate Professor and Head of the Department of Energy and Power Systems at the University of Zagreb Faculty of Electrical Engineering and Computing. He is also a leading researcher at the Innovation Centre Nikola Tesla (ICENT). His research interests include planning, operation, control, and economics of power and energy systems.

\end{IEEEbiography}
\end{comment}

\end{document}

%% file: appendix3.tex
%\begin{strip}
\section*{Appendix}\label{sec:App}
%\end{strip}
\subsection{Parameters}\label{sub:param}

The following parameters are defined over parameters from the nomenclature and are used to shorten the model
formulation. Parameters defined in \eqref{eq:P.1.1}--\eqref{eq:P.2.2} are used in the lower-level primal problem for power flow constraints \eqref{eq:LL.4}--\eqref{eq:LL.7} and subsequently in the dual problem. The naming scheme is inspired by the parameters definition, i.e. \emph{cps} stands for \emph{cosine-plus-sine} and \emph{cms} stands for \emph{cosine-minus-sine}. The other three parameters defined in \eqref{eq:P.3}--\eqref{eq:P.5} are used to shorten the SOC constraint reformulation from Subsection \ref{sub:reformulation} and, subsequently, in the dual problem.

\begin{equation}\tag{A.1.1}
\begin{split}
\cospsinij \!:=\! \gl \!\cdot\! \mathrm{cos}(\fitn\!-\!\fitm\!-\!\shift)&\!+\!\bl\! \cdot\! \mathrm{sin}(\fitn\!-\!\fitm\!-\!\shift),\\
&\!\!\quad \forall{t,(e,i,j)\in \AF}\\[-0.95cm]
\label{eq:P.1.1}
\end{split}
\end{equation}

%\hfill

\begin{equation}\tag{A.1.2}
\begin{split}
\cospsinij \!:=\! \gl \!\cdot\! \mathrm{cos}(\fitn\!-\!\fitm\!+\!\shift)&\!+\!\bl\! \cdot\! \mathrm{sin}(\fitn\!-\!\fitm\!+\!\shift),\\
&\!\!\quad \forall{t,(e,i,j)\in \AT}\\[-0.95cm]
\label{eq:P.1.2}
\end{split}
\end{equation}

\begin{equation}\tag{A.2.1}
\begin{split}
\cosmsinij \!\hspace{-0.95pt}:=\! \hspace{-0.7pt}\bl\! \hspace{-1.5pt} \cdot \!\mathrm{cos}(\fitn\!-\!\fitm\!-\!\shift)&\!-\!\gl \hspace{-0.65pt} \!\cdot\! \mathrm{sin}(\fitn\!-\!\fitm\!-\!\shift),\\
&\!\!\quad \hspace{-1pt} \forall{t,(e,i,j)\in \AF}\\[-0.95cm]
\label{eq:P.2.1}
\end{split}
\end{equation}

\begin{equation}\tag{A.2.2}
\begin{split}
\cosmsinij \! \hspace{-0.95pt} :=\!\hspace{-0.7pt} \bl\! \hspace{-1.5pt} \cdot \!\mathrm{cos}(\fitn\!-\!\fitm\!+\!\shift)&\!-\!\gl \hspace{-0.65pt} \!\cdot\! \mathrm{sin}(\fitn\!-\!\fitm\!+\!\shift),\\
&\!\!\quad \hspace{-1pt} \forall{t,(e,i,j)\in \AT}\\[-0.95cm]
\label{eq:P.2.2}
\end{split}
\end{equation}

\vspace{2pt}

\begin{equation}\tag{A.3}
\!\vpJedan \!:= \! \frac{-\gl\!\cdot\! \mathrm{cos} (\fitn\!-\!\fitm\!-\!\shift) }{\sqrt{\gl+\gt}\cdot \tap}, \!\!\!\quad \forall{t,(e,i,j)\in \AF}
\label{eq:P.3}
\end{equation}

\begin{equation}\tag{A.4}
\vpDva \!:= \! \sqrt{\gl+\gf}, \quad \forall{t,e}
\label{eq:P.4}
\end{equation}

\hfill

\vspace{0pt}

\begin{equation}\tag{A.5}
\begin{split}
&\vpTri \!:= \! \hspace{-0.5pt} \sqrt{\frac{\gl^2 \!\cdot \! \mathrm{sin}^2(\fitn\!\!-\!\fitm\!\!-\!\shift) \hspace{-0.5pt} \!+\!\! \gl \hspace{-1pt} \!\cdot\! (\gf\!\!+\!\gt) \!+\! \gf \!\cdot\! \gt}{\gl\!+\!\gf}},\\
&\hspace{157pt}\forall{t,(e,i,j)\in \AF}\\[-0.95cm]
\label{eq:P.5}
\end{split}
\end{equation}

\vspace{-16pt}

\subsection{Dual}\label{sub:dual}

The dual objective function \eqref{eq:D.1} is followed by constraints \eqref{eq:D.2}--\eqref{eq:D.9}, obtained by derivation of the primal problem in variable order as appearing in the $\Xill$ set from Subsection \ref{sub:initial}. The next three constraints \eqref{eq:D.10}--\eqref{eq:D.12} are the dual SOC constraints of \eqref{eq:SOC.1.1}, \eqref{eq:SOC.2.1} and \eqref{eq:LL.12}, respectively (due to the self-duality principle of the basic cones). The last constraint \eqref{eq:D.13} is a nonnegativity condition for the dual variables associated to the primal inequality constraints.

\begin{equation}\tag{B.4}
\begin{split}
& - \! \lambdaJedan \! \hspace{-1pt}+\!\! \underset{:(e,i,j)\in \AF}{\lambdaTri} \! \hspace{-1pt}+\hspace{-1pt} \!\!\! \underset{:(e,i,j)\in \AT}{\lambdaCetiri} \!\!\!\hspace{-0.6pt}-\hspace{-0.6pt} \! \underset{:\conds}{\lambdaJedanaest} \! \hspace{-1pt} =\! \hspace{-1pt} 0, \!\!\!\!\quad \forall{t,\!(e,i,j)\!\hspace{-1pt}\in\!\hspace{-1.2pt} \AF \!\hspace{-1pt}\cup\! \hspace{-1pt}\AT}
\label{eq:D.4}
\end{split}
\end{equation}

\begin{equation}\tag{B.5}
\begin{split}
& - \! \hspace{-1pt}\lambdaDva \! \hspace{-1pt}+\!\! \underset{:(e,i,j)\in \AF}{\lambdaPet} \! \hspace{-1pt}+\hspace{-1pt} \!\!\! \underset{:(e,i,j)\in \AT}{\lambdaSest} \!\!\!\hspace{-0.6pt}-\hspace{-0.6pt} \! \underset{:\conds}{\lambdaDvanaest} \! \hspace{-1pt} =\! \hspace{-1pt} 0, \!\!\!\!\quad \forall{t,\!(e,i,j)\!\hspace{-1pt}\in\!\hspace{-1.2pt} \AF \!\hspace{-1pt}\cup\! \hspace{-1pt}\AT}
\label{eq:D.5}
\end{split}
\end{equation}

\vspace{-6pt}

\begin{equation}\tag{B.6}
\begin{split}
& \C + \muJedanUp - \muJedanDn \!+ \!\! \sum_{\substack{i \\ :k \in \BG}}\lambdaJedan \!=\! 0, \quad \forall{t,k} :\CC=0\\[-0.98cm]
\label{eq:D.6}
\end{split}
\end{equation}

\vspace{4pt}

\vspace{-12pt}

\begin{equation}\tag{B.7}
\begin{split}
&  \muDvaUp - \muDvaDn \!+ \!\! \sum_{\substack{i \\ :k \in \BG}}\lambdaDva \!=\! 0, \quad \forall{t,k} \\[-0.98cm]
\label{eq:D.7}
\end{split}
\end{equation}

\vspace{0pt}

\begin{equation}\tag{B.8}
\begin{split}
& \underset{:\condc}{\lambdaSesnaest} \!-\!\hspace{-0.75pt} \underset{:\condc}{\lambdaPetnaest} \! +\! \hspace{-0.9pt} \underset{:\lnot\condc}{\lambdaDvadesetdva}  \!\hspace{-0.9pt} +\! \hspace{-1pt} \Vtn  \!\! \cdot\!  \Vtm  \!\cdot \hspace{-13pt} \sum_{(e,i,j) \in \AF}\hspace{-6pt} \!\!(\cospsinij \! \cdot \! \lambdaTri \\
&+\! \cospsinji \!\cdot \! \lambdaCetiriji \!-\!\cosmsinij \! \cdot \!\lambdaPet \!\\[3pt]
&-\! \cosmsinji \! \cdot \! \lambdaSestji  )/\tap \!=\! 0, \quad \forall{t,(i,j)\in \BP}\\[-0.95cm]
\label{eq:D.8}
\end{split}
\end{equation}

\vspace{0pt}

\begin{equation}\tag{B.9}
\begin{split}
(-\lambdaTri\!-\!\lambdaCetiriji \!+\! \underset{:\condv}{\lambdaSedamnaest} \!-\!& \underset{:\condv}{\lambdaDvadeset})/2 \!+\! \underset{:\lnot \condv}{\lambdaDvadesetjedan}\!=\!0,\\
&\quad \forall{t,(e,i,j)\in \AF}\\[-0.95cm]
\label{eq:D.9}
\end{split}
\end{equation}

\begin{equation}\tag{B.10}
\begin{split}
\!\!\!\lambdaSedamnaest^2 \!+\! \lambdaOsamnaest^2 \!+\! \lambdaDevetnaest^2& \! \le \! \lambdaDvadeset^2,\\
& \!\!\forall{t,(e,i,j)\in \AF :\condv}
\label{eq:D.10}
\end{split}
\end{equation}

\begin{equation}\tag{B.11}
\!\lambdaCetrnaest^2\!+\! \lambdaPetnaest^2 \!\le\! \lambdaSesnaest^2, \!\!\!\quad \forall{t,(i,j)\in \BP} :\condc
\label{eq:D.11}
\end{equation}

\begin{equation}\tag{B.12}
\lambdaJedanaest^2\!\!+\!\! \lambdaDvanaest^2 \!\!\le\! \lambdaTrinaest^2,\!\!\!\!\quad \forall{t,\!(e,i,j)\!\in\!\! \AF \!:\!\conds}\!\!\!
\label{eq:D.12}
\end{equation}

\begin{equation}\tag{B.13}
\mu \ge 0
\label{eq:D.13}
\end{equation}

\vspace{-34pt}

\begin{strip}
\vspace{0pt}
\hrulefill
\vspace{0pt}
%\hrulefill

\begin{equation}\tag{B.1}
\begin{split}
&\underset{\Xidu}{\mathrm{Max}}\:\:\Omegad :=\sum_{t,k}  ( \cc  \!-\!  \Pmax  \!\cdot\!  \muJedanUp  \!+\!  \Pmin  \!\cdot\!  \muJedanDn  \!-\! \Qmax  \!\cdot\! \muDvaUp  \!+\!   \Qmin \!\cdot\!  \muDvaDn )  - \hspace{-4pt} \sum_{\substack{t,k \\ :\CC >0}} \hspace{-4pt} (\C \!+\! \muJedanUp \!-\! \muJedanDn \!+\! \hspace{-5pt} \sum_{  \substack{i \\ :k \in \BG}} \hspace{-4pt}\lambdaJedan)^2/(4 \!\cdot\! \CC)\\[0pt]
&+\!\sum_{t,i} \!\left( (\Vtn \!-\! \Vmax)\cdot \muTriUp + (\Vmin \!-\! \Vtn) \cdot \muTriDn \right )\!-\!\!\sum_{t,i}\!\!\sum_{\;l\in \BL} \!\!(\dP \!\cdot\! \lambdaJedan  \!+\!  \dQ  \!\cdot\! \lambdaDva) -\!\!\sum_{t,i}\!\!\sum_{\;s\in \BS} \!\!(\gs \!\cdot\! \lambdaJedan \!-\! \bs  \!\cdot\! \lambdaDva)\! -\hspace{-15pt}\sum_{\substack{t,(i,j) \in \BP \\ :\lnot \condc}} \hspace{-13pt}  \lambdaDvadesetdva\\[0pt]
&+\hspace{-14pt} \sum_{\substack{t,(i,j)\in \BP \\ :\condc}}\hspace{-11pt}(3/4 \!\cdot\! \lambdaPetnaest \!-\! 5/4 \!\cdot\! \lambdaSesnaest) \!-\hspace{-13pt}\sum_{\substack{t,(e,i,j) \in \AF \\ :\condv}} \hspace{-11pt} (\lambdaSedamnaest  \!+\! \lambdaDvadeset)/2 \!+\hspace{-5pt} \sum_{t,i \in \RB} \hspace{-4pt}\fitn \!\cdot\! \lambdaSedam \!- \hspace{-4.5pt} \sum_{t,i\in \BB} \hspace{-3pt} \qbat  \!\cdot  \! \lambdaJedan \!-\hspace{-4.5pt}  \sum_{t,i\in \BB} \hspace{-3pt} \qqbat  \!\cdot  \! \lambdaDva \!- \hspace{-22.5pt} \sum_{\substack{t,(e,i,j) \in \A \\ :\conds}} \hspace{-20pt} \Sline \hspace{-0.5pt} \!\cdot \!  \lambdaTrinaest\\[0pt]
&-\hspace{-10pt}\sum_{t,(e,i,j)\in \AF} \hspace{-13pt} \left ((\gl \!+\!\gf) \!\cdot\! \lambdaTri \!-\! (\bl\!+\!\bf) \!\cdot\! \lambdaPet \right) \!\cdot\! (\Vtn)^2/\tap^2-\hspace{-14pt}\sum_{t,(e,i,j)\in \AT} \hspace{-16pt} \left ((\gl \!\!+\!\gt) \!\cdot\! \lambdaCetiri \!-\! (\bl \!+\! \bt)  \!\cdot\!  \lambdaSest \right) \!\cdot\! (\Vtn)^2\\[-1.33cm]
\label{eq:D.1}
\end{split}
\end{equation}

%\end{strip}
\vspace{10pt}

%\begin{strip}

\begin{equation}\tag{B.2}
\begin{split}
&\Vtn  \!\cdot\!\!\!\!\!\! \sum_{(e,i,j) \in \AF}\!\!\!\!\!\Vtm \cdot (\cosmsinij\!\cdot\! \lambdaTri \!-\! \cosmsinji\! \cdot \!\lambdaCetiriji \!+\!\cospsinij \!\cdot\! \lambdaPet \!-\! \cospsinji \!\cdot \! \lambdaSestji)/\tap\\[0pt]
&-\!\Vtn  \!\cdot\!\!\!\!\!\!\!\! \sum_{(e,i,j) \in \AT}\!\!\!\!\!\!\Vtm \!\cdot \! (\cosmsinji\!\cdot\! \lambdaTriji \!-\! \cosmsinij\! \cdot \!\lambdaCetiri \!+\!\cospsinji \!\cdot\! \lambdaPetji \!-\! \cospsinij \!\cdot \! \lambdaSest)/\tap\\[0pt]
&+\!\underset{:i\in \RB}{\lambdaSedam}\;-\hspace{-10pt}\sum_{\substack{(i,j)\in \BP\\ :\condc}}\hspace{-10pt} \lambdaCetrnaest/\sqrt{2} + \hspace{-12pt} \sum_{\substack{(i,j)\in \BPR\\ :\condcji}}\hspace{-10pt}\lambdaCetrnaestji/\sqrt{2} \!=\! 0, \quad \forall{t,i}
\label{eq:D.2}
\end{split}
\end{equation}

\end{strip}
\vspace{0pt}
\begin{strip}

\vspace{-30pt}
\begin{equation}\tag{B.3}
\begin{split}
&\muTriUp \!-\! \muTriDn \!+\! 2\!\cdot\! \Vtn \!\cdot  (\lambdaDva \!\cdot \!\!\! \sum_{s\in\BS}\!\bs \!-\! \lambdaJedan \!\cdot\!\!\!\sum_{s\in\BS}\!\gs) +\hspace{-11pt}\sum_{\substack{(e,i,j)\in \AT \\ :\condv}}\hspace{-10pt} \vpDva \! \cdot \! \lambdaOsamnaestji - \hspace{-10pt}\sum_{\substack{(e,i,j)\in \AF \\ :\condv}}\hspace{-7pt} (\vpJedan \!\cdot \!  \lambdaOsamnaest \!+\! \vpTri \!\cdot\! \lambdaDevetnaest)\\
&+\hspace{-10pt}\sum_{(e,i,j)\in \AF}\hspace{-9pt} \left[\:\left(-2\!\cdot\!(\gl\!+\!\gf)\!\cdot\! \Vtn/\tap^2 \! + \! \Vtm \!\!\cdot\! \cospsinij/\tap\right)\right. \!\! \cdot \! \lambdaTri + \left(2 \! \cdot \! (\bl \!+\! \bf) \!\cdot\! \Vtn/\tap^2 \!-\! \Vtm \!\! \cdot \! \cosmsinij/\tap \right) \!\cdot\! \lambdaPet\\
&\left.+\Vtm \! \cdot  (\cospsinji \! \cdot \! \lambdaCetiriji - \cosmsinji \cdot \lambdaSestji) / \tap \:\right]\\[2pt]
&+\hspace{-12pt}\sum_{(e,i,j)\in \AT}\hspace{-12pt} \left[\:\left(-2\!\cdot\!(\gl\!+\!\gt)\!\cdot\! \Vtn \! + \! \Vtm \!\!\cdot\! \cospsinij/\tap\right)\right.\! \! \cdot \! \lambdaCetiri + \left(2 \! \cdot \! (\bl \!+\! \bt) \!\cdot\! \Vtn \!-\! \Vtm \!\! \cdot \! \cosmsinij/\tap \right) \!\cdot\! \lambdaSest\\
&\left.+\Vtm \! \cdot  (\cospsinji \! \cdot \! \lambdaTriji - \cosmsinji \cdot \lambdaPetji) / \tap \:\right] \!=\! 0, \quad \forall{t,i}\\[-0.95cm]
\label{eq:D.3}
\end{split}
\end{equation}

\hrulefill
\end{strip}

Set of dual variables $\Xidu$ contains all $\lambda$ and $\mu$ variables.
%\vspace{7pt}

%\vspace{6pt}